\input{aipcheck}

\documentclass[
    ,final            
  ]
  {aipproc}

\layoutstyle{6x9}

\usepackage{graphicx} 
\usepackage{wasysym}

\def\aj{AJ}
\def\araa{ARA\&A}
\def\apj{ApJ}

\def\apjs{ApJS}

\def\aap{A\&A}

\def\aaps{A\&AS}
\def\mnras{MNRAS}

\def\angstrom {\AA}
\def\Teff {T$_{\rm eff}$}
\def\logg {log $g$}

\begin{document}
\title{Spectral libraries and their uncertainties}


\author{P. Coelho}{
  address={Institut d'Astrophysique, 98 bis Bd Arago, 75014 Paris, France}
}

\begin{abstract}
Libraries of stellar spectra are fundamental tools in the study of stellar populations and 
in automatic determination of atmospheric parameters for large samples of observed stars. In the 
context of the present volume, here I give an overview of the current status of stellar spectral libraries 
from the perspective of stellar population modeling: what we have currently available, how good 
they are, and where we need further improvement.
\end{abstract}

\maketitle


\section{Introduction}

An evolutionary stellar population model [e.g. 1] has two main ingredients: a set of stellar 
evolutionary models (tracks or isochrones) that predicts how the stars are distributed 
in the HR diagram, and a library of stellar observables (e.g. colours, spectra, spectral 
indices) that is used to predict the colours or spectra of a stellar population, given the 
evolutionary predictions (see articles by A. Vazdekis and S. Cassissi in this volume). 

This article focus on the library of stellar observables, more specifically, on libraries 
of stellar spectra. A good stellar library is at the very heart of accurate SP models, and 
should ideally provide complete coverage of the HR diagram, accurate atmospheric parameters 
(effective temperature \Teff, surface gravities \logg, metallicities Z and abundances 
[Fe/H], [Mg/Fe], etc.), and good wavelength coverage and/or good spectral resolution, 
depending on the aimed application. Either empirical or theoretical libraries can 
be used in stellar population (SP) modeling and both types have improved dramatically 
in recent years, allowing the construction of more accurate SP models. There are plenty 
of models in literature using both empirical (e.g. [2, 3], the visible range of [4]) and 
theoretical libraries (e.g. [5, 6]). Observations are also becoming increasingly better in 
terms of spectral resolution and coverage and demanding more from the modelling point 
of view [e.g. 7]. 

A good starting point to inspect the libraries available in literature is the comprehensive list maintained by 
D. Montes\footnote{\url{http://www.ucm.es/info/Astrof/invest/actividad/spectra.html}}, 
with more than 80 libraries currently listed. The 
libraries cover a virtually complete wavelength range with theoretical libraries, and from 
near UV to K-band with empirical ones, at spectral resolutions $R = \Delta\lambda/\lambda$ from 200 to 
80000 (even higher for a few selected stars), and different coverages of atmospheric 
parameters and abundances. In the following sections I highlight the strengths and caveats 
of empirical and theoretical libraries currently available.

\section{EMPIRICAL LIBRARIES}

An empirical library is an homogeneous compilation of observed stellar spectra. It is 
not a simple task to assemble a library that simultaneously features high S/N, good 
flux calibration, large wavelength coverage, high spectral resolution and accurately 
derived stellar parameters. Major improvements have been made in the last years, with 
the publications of empirical libraries with improved spectral resolution and parameter 
coverage: e.g. STELIB [8], UVES POP [9], Indo-US [10], ELODIE [11, 12], MILES 
[13, 14]. Being based on real stars, the major advantage of an empirical library is that 
the spectral properties are highly reliable, limited only by the quality of the observations. 
High quality observations are limited to the closest stars, and thus the coverage of the 
HR diagram and abundances are biased towards the typical stellar population targeted 
by the observations. 

The coverage of two of the most complete libraries available nowadays, ELODIE and 
MILES, are shown in Fig. 1 in \Teff~vs. \logg~space. 
ELODIE\footnote{\url{http://www.obs.u-bordeaux1.fr/m2a/soubiran/elodie_library.html}} in its current version 
(3.1) contains 1388 starts, covers the wavelength range from 4000 to 6800\angstrom, and has a 
typical S/N of 150 per pixel at 5550\angstrom. Although it has a somewhat limited wavelength 
coverage, it has very high spectral resolution (R = 10000 for flux calibrated spectra and 
R = 42000 for flux normalised to the pseudo-continuum). 
MILES\footnote{\url{http://www.ucm.es/info/Astrof/miles/miles.html}} was assembled trying 
to fill major gaps that existed in previous empirical libraries in terms of HR coverage. It 
has 985 stars with spectra ranging from 3525 to 7500\angstrom~at a 2.3\angstrom~(FWHM) resolution. 
Stars from ELODIE and MILES are shown in the upper and lower panels in Fig. 1 
respectively, for three bins of [Fe/H] as indicated. Isochrones from [15] for ages 30Myr, 
100Myr, 1Gyr and 10Gyr are overplotted. Clearly the bin with the iron abundance 
around the solar value is the most complete. A possible caveat in this regime is that 
the most luminous giants in the upper part of the red giant branch and asymptotic giant 
branch evolutionary phases are not yet being covered, but this should not have a large 
effect modeling the visible wavelength range. Empirical libraries that focus specifically 
on those stars [e.g. 16] may be used to cover these phases if needed. Outside the solar 
metallicity regime, the coverage is less complete: in the super-solar regime, populations 
older than $\sim$ 100 Myr can still be modeled reliably, and in the case of metal poor 
populations ([Fe/H] $\sim$ -1.8), only old populations ($\sim$ 10 Gyr) can be modeled. 

The main caveat of empirical libraries is that SP models based solely on them are 
not 
able to reproduce consistently the medium to high resolution spectral features of systems 
which have undergone a star formation history different than the solar-neighborhood. 
The first compelling evidence of this limitation was presented by [17] (see also article 
by R. Peletier in this volume), who showed that SP models for Lick/IDS indices cannot 
reproduce the indices measured in elliptical galaxies, indicating that these systems are 
overabundant in $\alpha$-elements relative to the Sun. This happens because, by construction, 
the abundance pattern of models based on empirical libraries is dictated by that of the 
library stars, which mirrors the abundance pattern mainly of the solar neighborhood 
[e.g. 18]. Theoretical libraries must be used to overcome this limitation, either solely or
in combination with empirical libraries [see e.g. 19].

\begin{figure}
  \includegraphics[width=\textwidth]{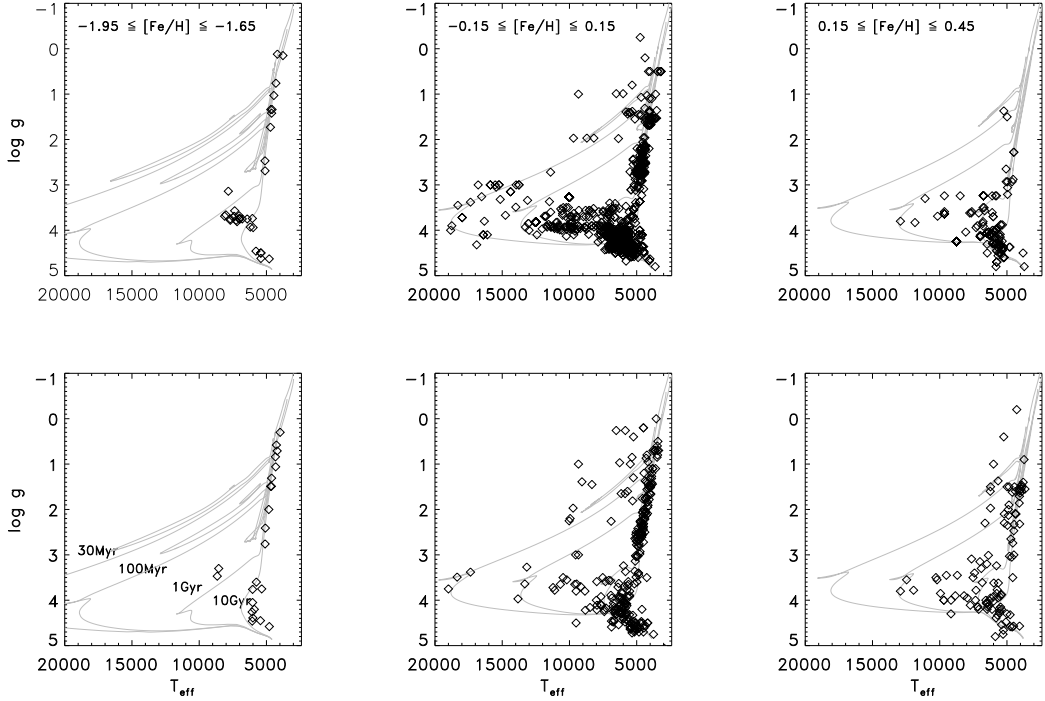}
  \caption{Coverage of ELODIE and MILES stellar libraries (top and bottom rows, respectively) in 
\Teff~vs. \logg space. Three bins of [Fe/H] are shown, as indicated in the top figures. Isochrones from [15] 
are overplotted, for ages 30 Myr, 100 Myr, 1 Gyr and 10 Gyr. The abundance [Fe/H] of the isochrones are 
-1.75, +0.06 and 0.25 dex for the left-hand, central and right-hand panels respectively.}
\end{figure}

\section{THEORETICAL LIBRARIES}

A theoretical (or synthetic) spectral library is based on model atmospheres predictions 
and atomic and molecular line lists. A model atmosphere is the run of temperature, 
pressure (gas, electron and radiation), convective velocity and flux, and more generally 
of all relevant quantities as a function of some depth variable (geometrical, optical depth 
at some special frequency, or column mass). The synthetic spectrum or flux distribution 
is the emergent flux computed based on a model atmosphere and atomic and molecular 
line opacity lists, and is required for comparison with observations. Theoretical libraries 
have the advantage of covering the parameter space in \Teff, \logg, and abundances at 
will. Moreover, a synthetic star has very well defined atmospheric parameters, infinite 
S/N, and covers larger wavelength ranges at higher resolutions than observed spectra. 
To compute a large synthetic library can be demanding in terms of computational time, 
but it is usually feasible. The caveat of theoretical libraries is that, being based on our 
knowledge of the physics of stellar atmospheres and databases of atomic and molecular 
transitions, those libraries are limited by the approximations and (in)accuracies of their
underlying models and input data. 

The ability of theoretical libraries in predicting broad-band colors and medium to high 
resolution spectral features has been assessed by e.g. [20, 21, 22, 23]. Studying libraries 
with large coverages in stellar atmospheric parameters, [22] finds that in general current 
models are able to reproduce stellar colors with accuracy for a fair interval in effective 
temperatures and gravities, but there are still some problems with $U-B$ and $B-V$ 
colors, and very cool stars in general (\Teff~$<$ 4000K). Studying models for red giants, 
[21] finds that theoretical \Teff~vs. color relations for visible and near-infrared colors 
agree with observations within 100K down to \Teff~$\sim$ 3400K, but none of the existing 
theoretical relations reproduce the data below $\sim$3800K in $U-B$ and $B-V$. 

Concerning medium and high resolution features, [22] analyzed the performance of 
recent high-resolution libraries by comparing their predictions of spectral indices (from 
3500 to 8700) to measurements from empirical libraries. Here also it is found that 
many indices are well reproduced for a fair range of temperatures, but that lists of atomic 
and molecular opacities still need improvement in the blue region of the spectrum and 
for the cool stars regime in general. This is illustrated in Fig. 2, where predictions of the 
theoretical library by Coelho et al. [24] are plotted against observed values from MILES 
library, for six spectral indices. In general the agreement is good, but deviations appear 
in the cool stars regime (as easily seen for the indices H$\gamma$, G4300 and Ca4455). Studying 
the performance of theoretical spectra in reproducing high resolution observations, [23] 
compared model predictions to three outstanding reference templates, namely the Sun, 
Arcturus and Vega. These authors fitted the high-resolution (R = 522000) spectrum of 
the Sun in the region 3500 to 7000\angstrom~and found that the observed solar flux is reproduced 
within 9\% in relative flux uncertainty (rms). Downgrading the spectrum to R = $10^5$ 
brings the rms down to 5\%. The agreement between model and observations is better 
for Vega (1\%), and worse for Arcturus (9\%) at the same R = $10^5$ resolution. They 
further compared the theoretical predictions with a selected sample of observations from 
ELODIE at R = 42000, and found that there is a trend in temperature and gravity: 
hotter stars are better reproduced than cooler ones, and dwarfs are better reproduced 
than giants. No trend with metallicity was found. 

The origin of the deficiencies in the theoretical predictions may be two-fold: the 
underlaying physics theory and/or the input physics data (atomic and molecular opacities) 
of the models. In terms of physics theory, most of the libraries computed to cover large 
parameter spaces are computed under 1D and LTE assumptions. For stars of spectral 
types from A to G these approximations are in general suitable, but for other spectral 
types it is known that effects of 3D hydrodynamics, N-LTE, winds and chromospheric 
contribution may be important. In the regime of cool stars effects such as convection, 
variability, mass loss and dust formation become increasingly important [e.g. 
25, 26, 27]. State-of-the-art hydrodynamical models are an important recent advancement
 [e.g. 28, 29], but these models are considerably time-consuming and until now no 
extensive grids of 3D models exist. In the hot stars regime, models taking into account 
N-LTE effects are mature and grids of N-LTE line-blanketed model atmospheres and 
fluxes of O- and B-type stars are available in literature [e.g. 30, 31, 32]. The modeling 
of mass loss and winds (particularly important in the UV wavelength range) is less well 
established (C. Leitherer, priv. comm.), but progress is being made [e.g. 33, 34]. 

In terms of physics basic data, databases of atomic and molecular transitions provide
fewer lines with highly accurate oscillator strengths and broadening parameters than we 
would like, besides being often incomplete [see e.g. 35]. Constant updates and improvements 
are necessary from laboratory data, critical compilations, empirical adjustments 
and improved quantum mechanics computations [e.g. 36, 37, 38]. Some libraries of
theoretical spectra opt for including atomic and molecular line lists which are as complete 
as possible, thus providing a good treatment of line-blanketing (i.e. include the so-called 
'predicted lines', see [39]). These libraries are good for spectrophotometric predictions 
and low-spectral resolution studies [e.g. 40, 41, 42, 43]. But most of the predicted lines 
does not have accurate wavelengths and oscillator strengths, and thus are not appropriate 
to reproduce high spectral resolution features. Therefore other theoretical libraries were 
computed with shorter, fine-tuned, empirically calibrated atomic and molecular line lists 
[e.g. 44, 45, 24]. These libraries are much more accurate in reproducing high-resolution 
features, but are not as accurate in predicting colors due to the missing line-blanketing. 
The inclusion or not of the predicted lines has a clear impact on the colors predicted 
by SP models as shown in Fig. 12 by [6]. In order to provide accurate predictions for 
both high spectral resolution features and broad-band colors, either libraries that do not 
include the predicted lines must be flux calibrated [e.g. section 3.2 in 6], or low and high
resolution SP models should be computed with different libraries [as adopted by 46].

\begin{figure}
  \includegraphics[width=\textwidth]{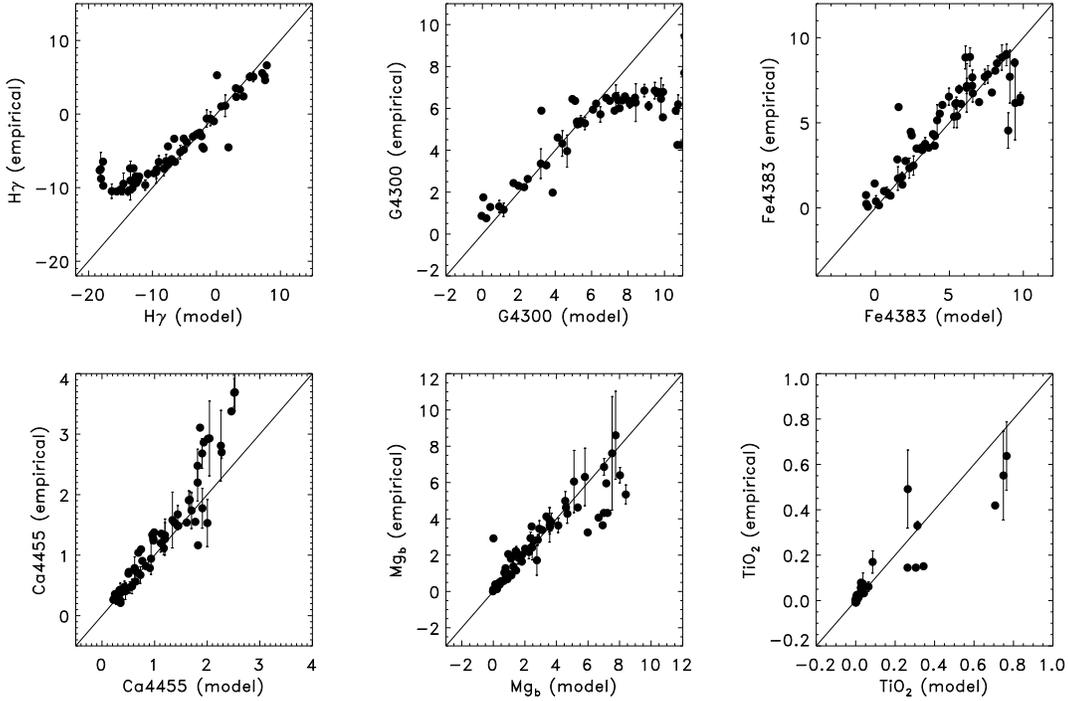}
  \caption{Comparison between predictions of the theoretical library by Coelho et al. in the x-axis and 
indices measured in the MILES empirical library in the y-axis, for six spectral indices indicated by the 
labels. Stars in the empirical libraries were binned in parameter space with $\Delta$\Teff = 250K and $\Delta$\logg = 0.5 
dex (reasonable values given the atmospheric parameters uncertainties). Mean values in each parameter 
bin were plotted against the theoretical predictions, and error bars illustrate the 1 sigma deviations. Points 
without error bars occur when only one star in the empirical library falls into a parameter bin. The solid 
line is the one-to-one relation. }
\end{figure}

\section{A NOTE ON ATMOSPHERIC PARAMETERS}
Accurate atmospheric parameters $-$ \Teff, \logg~and abundances $-$ are essential to link the 
stars from the spectral library to stellar evolution predictions, the other crucial ingredient 
of a SP model. In one hand, quite often the parameters of observed stellar spectra 
are derived by comparison to models, or to calibrations which are largely based on 
models [e.g. 20]. On the other hand, modelers of stellar spectra need stars with \Teff~and 
\logg~derived by fundamental ways (independent or weakly dependent on models, see 
e.g. [47]) in order to test and calibrate the models. In the case of temperatures, for 
example, direct estimation of \Teff~is possible for close stars if the angular diameter of a 
star is known (interferometric measurements or lunar occultations) [e.g. 48, 49]. Recent 
determinations are able to determine \Teff~with a typical accuracy of 5\% (better for well 
studied stars such as Arcturus, see e.g. compilations in [50, 21]), placing a lower limit 
in the absolute accuracy we can obtain for \Teff. Moreover, bellow $\sim$ 3400K most (if not 
all) giants are variable, and we may wonder what published values of \Teff~and \logg~for 
those stars really mean. 

In the case of empirical libraries, methods of deriving the atmospheric parameters 
based on a reference sample of well studied stars [e.g. TGMET: 51, 52] are crucial to 
guarantee homogeneous estimations, and were indeed adopted by e.g. [11, 14]. It does 
not guarantee, however, against systematic errors, if the parameters of the reference 
stars are affected by undetected systematic deviations. Moreover, the reference stars 
usually encompass a limited range of spectral types, and outside this range the derived 
parameters are less reliable. In the case of theoretical libraries atmospheric parameters 
are known by construction, but evidence show that models have room to improve yet, as 
discussed previously. In this context, an exercise done in [23] in very appropriate: these 
authors compared the high resolution spectrum of the Sun to a small grid of theoretical 
libraries, and derived the solar parameters using the theoretical grid as reference stars. 
The parameters derived for the Sun have offsets with respect to the real values of 
$\Delta$\Teff~= +80K, $\Delta$\logg~= +0.5 and $\Delta$[Fe/H]= -0.3. In a sense, these offsets quantify 
the accuracy of the theoretical libraries in a scale that can be directly compared to the 
uncertainties of the atmospheric parameters in empirical libraries. 

In any case, the impact of (realistic) errors of the atmospheric parameters on the 
predictions of SP models remains to be quantified.

\section{CONCLUSIONS}
Empirical stellar libraries are mature, and are the most reliable for the modeling of SP in 
the optical and near-IR wavelengths. The coverage in terms of \Teff~and \logg~is very good 
around solar metallicities. Outside the solar metaliticy regime, the range of SP ages that 
can be modeled is somewhat restricted. The main caveat of empirical libraries is their 
inability to model accurately populations that have undergone a star formation history 
different than the stars in the library (biased towards the solar neighborhood). This is
an issue in interpreting large samples of medium to high spectral resolution integrated 
spectra of clusters and galaxies. Future advancements in empirical libraries will likely 
come from space telescopes, to better explore regions beyond the visible range [e.g. 
NGSL by 53]. 

Theoretical libraries are to some extend well calibrated in the visible and near-IR for 
stars as late as G-type. Colors in the visible and near-IR bands are reproduced within the 
error bars for temperatures down to $\sim$ 3500K. At a resolution of R = $10^5$ , the spectrum 
of the Sun is today reproduced in 5\% of relative flux, Arcturus is reproduced in 9\% and 
Vega is reproduced in 1\%. Residuals are in general larger towards cooler stars or lower 
surface gravities. For stars below $\sim$ 3500K, current developments in hydrodynamical 
models and pulsating atmospheres should improve the accuracy of the models, but it 
may well take some years before such grids are available to the completeness needed 
for population synthesis. For O- and B-type stars, recent developments in mass loss 
modeling, expanding atmospheres and wind features are being incorporated into the 
theoretical grids to model the UV with better accuracy. 

Theoretical (or possibly semi-theoretical) libraries are the most promising to model 
the integrated spectral features of populations beyond the local one. As the observations 
of extra-galactic populations improve in terms of spectral resolution, environment and 
redshift coverage, it is my personal view that next generations of libraries for SP 
modeling will tend more and more towards theoretical libraries, with empirical libraries being 
used for the crucial testing and calibration of the theoretical spectra. 

\emph{Acknowledgments:} The author acknowledges the support of the European Community 
under a Marie Curie International Incoming Fellowship (6th Framework Programme, 
FP6).


\begin{thebibliography}{53}
\expandafter\ifx\csname natexlab\endcsname\relax\def\natexlab#1{#1}\fi
\providecommand{\enquote}[1]{``#1''}
\expandafter\ifx\csname url\endcsname\relax
  \def\url#1{\texttt{#1}}\fi
\expandafter\ifx\csname urlprefix\endcsname\relax\def\urlprefix{URL }\fi
\providecommand{\eprint}[2][]{\url{#2}}

\bibitem[{Tinsley}(1980)]{tinsley80}
B.~M. {Tinsley}, \emph{Fundamentals of Cosmic Physics} \textbf{5}, 287--388
  (1980).

\bibitem[{Vazdekis}(1999)]{vazdekis99}
A.~{Vazdekis}, \emph{ApJ} \textbf{513}, 224--241 (1999).

\bibitem[{Le Borgne} et~al.(2004)]{PEGASE-HR}
D.~{Le Borgne}, B.~{Rocca-Volmerange}, P.~{Prugniel}, A.~{Lan{\c c}on},
  M.~{Fioc}, and C.~{Soubiran}, \emph{A\&A} \textbf{425}, 881--897 (2004).

\bibitem[{Bruzual} and {Charlot}(2003)]{BC03}
G.~{Bruzual}, and S.~{Charlot}, \emph{MNRAS} \textbf{344}, 1000--1028 (2003).

\bibitem[{Delgado} et~al.(2005)]{delgado+05}
R.~M.~G. {Delgado}, M.~{Cervi{\~n}o}, L.~P. {Martins}, C.~{Leitherer}, and
  P.~H. {Hauschildt}, \emph{MNRAS} \textbf{357}, 945--960 (2005).

\bibitem[{Coelho} et~al.(2007)]{coelho+07}
P.~{Coelho}, G.~{Bruzual}, S.~{Charlot}, A.~{Weiss}, B.~{Barbuy}, and J.~W.
  {Ferguson}, \emph{\mnras} \textbf{382}, 498--514 (2007).

\bibitem[{Adelman-McCarthy} et~al.(2008)]{sdss-dr6}
J.~K. {Adelman-McCarthy}, M.~A. {Ag{\"u}eros}, S.~S. {Allam}, and {et al.},
  \emph{\apjs} \textbf{175}, 297--313 (2008).

\bibitem[{Le Borgne} et~al.(2003)]{stelib}
J.-F. {Le Borgne}, G.~{Bruzual}, R.~{Pell{\'o}}, A.~{Lan{\c c}on},
  B.~{Rocca-Volmerange}, B.~{Sanahuja}, D.~{Schaerer}, C.~{Soubiran}, and
  R.~{V{\'{\i}}lchez-G{\'o}mez}, \emph{A\&A} \textbf{402}, 433--442 (2003).

\bibitem[{Jehin} et~al.(2005)]{uvespop}
E.~{Jehin}, S.~{Bagnulo}, C.~{Melo}, C.~{Ledoux}, and R.~{Cabanac},
  \enquote{{The UVES Paranal Observatory Project: a public library of high
  resolution stellar spectra},} in \emph{IAU Symposium}, edited by V.~{Hill},
  P.~{Fran{\c c}ois}, and F.~{Primas}, 2005, pp. 261--262.

\bibitem[{Valdes} et~al.(2004)]{Indo-US}
F.~{Valdes}, R.~{Gupta}, J.~A. {Rose}, H.~P. {Singh}, and D.~J. {Bell},
  \emph{\apjs} \textbf{152}, 251--259 (2004).

\bibitem[{Prugniel} and {Soubiran}(2001)]{ELODIE}
P.~{Prugniel}, and C.~{Soubiran}, \emph{\aap} \textbf{369}, 1048--1057 (2001).

\bibitem[{Prugniel} et~al.(2007)]{ELODIE3}
P.~{Prugniel}, C.~{Soubiran}, M.~{Koleva}, and D.~{Le Borgne}, \emph{ArXiv
  Astrophysics}  (2007), \eprint{arXiv:astro-ph/0703658}.

\bibitem[{S{\'a}nchez-Bl{\'a}zquez} et~al.(2006)]{MILES1}
P.~{S{\'a}nchez-Bl{\'a}zquez}, R.~F. {Peletier}, J.~{Jim{\'e}nez-Vicente},
  N.~{Cardiel}, A.~J. {Cenarro}, J.~{Falc{\'o}n-Barroso}, J.~{Gorgas},
  S.~{Selam}, and A.~{Vazdekis}, \emph{\mnras} \textbf{371}, 703--718 (2006).

\bibitem[{Cenarro} et~al.(2007)]{MILES2}
A.~J. {Cenarro}, R.~F. {Peletier}, P.~{S{\'a}nchez-Bl{\'a}zquez}, S.~O.
  {Selam}, E.~{Toloba}, N.~{Cardiel}, J.~{Falc{\'o}n-Barroso}, J.~{Gorgas},
  J.~{Jim{\'e}nez-Vicente}, and A.~{Vazdekis}, \emph{\mnras} \textbf{374},
  664--690 (2007).

\bibitem[{Pietrinferni} et~al.(2004)]{pietrinferni+04}
A.~{Pietrinferni}, S.~{Cassisi}, M.~{Salaris}, and F.~{Castelli}, \emph{\apj}
  \textbf{612}, 168--190 (2004).

\bibitem[{Lan{\c c}on} and {Mouhcine}(2002)]{lancon_mouchine02}
A.~{Lan{\c c}on}, and M.~{Mouhcine}, \emph{\aap} \textbf{393}, 167--181 (2002).

\bibitem[{Worthey} et~al.(1992)]{worthey+92}
G.~{Worthey}, S.~M. {Faber}, and J.~J. {Gonzalez}, \emph{ApJ} \textbf{398},
  69--73 (1992).

\bibitem[{McWilliam}(1997)]{mcwilliam97}
A.~{McWilliam}, \emph{\araa} \textbf{35}, 503--556 (1997).

\bibitem[{Coelho}(in press)]{coelho08proc}
P.~{Coelho}, \enquote{{Model stars for the modelling of galaxies:
  alpha-enhancement in stellar populations models},} in \emph{Proceedings of
  ``XII Reunion Latinoamericana (2007)''}, RevMexAA, in press,
  \eprint{arXiv:astro-ph/0802.2665}.

\bibitem[{Bessell} et~al.(1998)]{bessell+98}
M.~S. {Bessell}, F.~{Castelli}, and B.~{Plez}, \emph{\aap} \textbf{333},
  231--250 (1998).

\bibitem[{Ku{\v c}inskas} et~al.(2005)]{kucinskas+05}
A.~{Ku{\v c}inskas}, P.~H. {Hauschildt}, H.-G. {Ludwig}, I.~{Brott},
  V.~{Vansevi{\v c}ius}, L.~{Lindegren}, T.~{Tanab{\'e}}, and F.~{Allard},
  \emph{\aap} \textbf{442}, 281--308 (2005).

\bibitem[{Martins} and {Coelho}(2007)]{MC07}
L.~P. {Martins}, and P.~{Coelho}, \emph{\mnras} \textbf{381}, 1329--1346
  (2007).

\bibitem[{Bertone} et~al.(2008)]{bertone+08}
E.~{Bertone}, A.~{Buzzoni}, M.~{Ch{\'a}vez}, and L.~H.
  {Rodr{\'{\i}}guez-Merino}, \emph{\aap} \textbf{485}, 823--835 (2008).

\bibitem[Coelho et~al.(2005)]{coelho+05}
P.~Coelho, B.~Barbuy, J.~Melendez, R.~Schiavon, and B.~Castilho, \emph{A\&A}
  \textbf{443}, 735 (2005).

\bibitem[{Allard} et~al.(2003)]{allard+03proc}
F.~{Allard}, T.~{Guillot}, H.-G. {Ludwig}, P.~H. {Hauschildt}, A.~{Schweitzer},
  D.~R. {Alexander}, and J.~W. {Ferguson}, \enquote{{Model Atmospheres and
  Spectra: The Role of Dust},} in \emph{Brown Dwarfs}, edited by
  E.~{Mart{\'{\i}}n}, 2003, vol. 211 of \emph{IAU Symposium}, pp. 325.

\bibitem[{Collet} et~al.(2005)]{collet+05}
R.~{Collet}, M.~{Asplund}, and F.~{Th{\'e}venin}, \emph{\aap} \textbf{442},
  643--650 (2005).

\bibitem[{Lan{\c c}on} et~al.(2007)]{lancon+07}
A.~{Lan{\c c}on}, P.~H. {Hauschildt}, D.~{Ladjal}, and M.~{Mouhcine},
  \emph{\aap} \textbf{468}, 205--220 (2007).

\bibitem[{Muthsam} et~al.(2007)]{muthsam+07}
H.~J. {Muthsam}, B.~{L{\"o}w-Baselli}, C.~{Obertscheider}, M.~{Langer},
  P.~{Lenz}, and F.~{Kupka}, \emph{\mnras} \textbf{380}, 1335--1340 (2007).

\bibitem[{Freytag} and {H{\"o}fner}(2008)]{freytag_hofner08}
B.~{Freytag}, and S.~{H{\"o}fner}, \emph{\aap} \textbf{483}, 571--583 (2008).

\bibitem[{Lanz} and {Hubeny}(2003)]{lanz_hubeny03}
T.~{Lanz}, and I.~{Hubeny}, \emph{\apjs} \textbf{146}, 417--441 (2003).

\bibitem[{Lanz} and {Hubeny}(2007)]{lanz_hubeny07}
T.~{Lanz}, and I.~{Hubeny}, \emph{\apjs} \textbf{169}, 83--104 (2007).

\bibitem[{Martins} et~al.(2005)]{martins+05}
L. {Martins}, R. {Delgado}, C.~{Leitherer}, M.~{Cervi{\~n}o}, and
  P.~{Hauschildt}, \emph{MNRAS} \textbf{358}, 49--65 (2005).

\bibitem[{Pauldrach} et~al.(2001)]{pauldrach+01}
A.~W.~A. {Pauldrach}, T.~L. {Hoffmann}, and M.~{Lennon}, \emph{\aap}
  \textbf{375}, 161--195 (2001).

\bibitem[{van Marle} et~al.(2008)]{vanmarle+08}
A.~J. {van Marle}, S.~P. {Owocki}, and N.~J. {Shaviv}, \emph{\mnras}
  \textbf{389}, 1353--1359 (2008).

\bibitem[{Kurucz}(2006)]{kurucz06}
R.~L. {Kurucz}, \enquote{{Including all the Lines},} in \emph{Radiative
  Transfer and Applications to Very Large Telescopes}, edited by P.~{Stee},
  2006, vol.~18 of \emph{EAS Publications Series}, pp. 129--155.

\bibitem[{Mel{\'e}ndez} and {Barbuy}(1999)]{melendez_barbuy99}
J.~{Mel{\'e}ndez}, and B.~{Barbuy}, \emph{\apjs} \textbf{124}, 527--546 (1999).

\bibitem[{Fuhr} and {Wiese}(2006)]{fuhr_wiese06}
J. {Fuhr}, and W. {Wiese}, \emph{Journal of Physical and Chemical
  Reference Data} \textbf{35}, 1669--1809 (2006).

\bibitem[{Wiese} and {Fuhr}(2007)]{fuhr_wiese07}
W. {Wiese}, and J. {Fuhr}, \emph{Journal of Physical and Chemical
  Reference Data} \textbf{36}, 1287--1345 (2007).

\bibitem[{Kurucz}(1995)]{kurucz95proc}
R.~L. {Kurucz}, \enquote{{An Atomic and Molecular Data Bank for Stellar
  Spectroscopy},} in \emph{Laboratory and Astronomical High Resolution
  Spectra}, edited by A.~J. {Sauval}, R.~{Blomme}, and N.~{Grevesse}, 1995,
  vol.~81 of \emph{Astronomical Society of the Pacific Conference Series}, pp.
  583.

\bibitem[{Westera} et~al.(2002)]{BASEL3}
P.~{Westera}, T.~{Lejeune}, R.~{Buser}, F.~{Cuisinier}, and G.~{Bruzual},
  \emph{A\&A} \textbf{381}, 524--538 (2002).

\bibitem[Castelli and Kurucz(2003)]{ATLASODFNEW}
F.~Castelli, and R.~L. Kurucz, \enquote{New Grids of ATLAS9 Model Atmospheres,}
  in \emph{Modelling of Stellar Atmospheres}, IAU Symp 210, Astronomical
  Society of the Pacific, 2003, p. A20.

\bibitem[{Brott} and {Hauschildt}(2005)]{PHOENIX05}
I.~{Brott}, and P.~H. {Hauschildt}, \enquote{{A PHOENIX Model Atmosphere Grid
  for Gaia},} in \emph{The Three-Dimensional Universe with Gaia}, edited by
  C.~{Turon}, K.~S. {O'Flaherty}, and M.~A.~C. {Perryman}, 2005, vol. 576 of
  \emph{ESA Special Publication}, p. 565.

\bibitem[{Gustafsson} et~al.(2008)]{MARCS08}
B.~{Gustafsson}, B.~{Edvardsson}, K.~{Eriksson}, U.~G. {J{\o}rgensen},
  {\AA}.~{Nordlund}, and B.~{Plez}, \emph{\aap} \textbf{486}, 951--970 (2008).

\bibitem[{Peterson} et~al.(2001)]{peterson+01}
R.~C. {Peterson}, B.~{Dorman}, and R.~T. {Rood}, \emph{ApJ} \textbf{559},
  372--387 (2001).

\bibitem[{Rodr{\'{\i}}guez-Merino} et~al.(2005)]{uvblue05}
L.~H. {Rodr{\'{\i}}guez-Merino}, M.~{Chavez}, E.~{Bertone}, and A.~{Buzzoni},
  \emph{ApJ} \textbf{626}, 411--424 (2005).

\bibitem[{Percival} et~al.(2009)]{percival+09}
S.~M. {Percival}, M.~{Salaris}, S.~{Cassisi}, and A.~{Pietrinferni},
  \emph{\apj} \textbf{690}, 427--439 (2009).

\bibitem[{Cayrel}(2002)]{cayrel02proc}
R.~{Cayrel}, \enquote{{Determination of Fundamental Parameters},} in
  \emph{Observed HR Diagrams and Stellar Evolution}, edited by T.~{Lejeune},
  and J.~{Fernandes}, 2002, vol. 274 of \emph{Astronomical Society of the
  Pacific Conference Series}, pp. 133.

\bibitem[{Code} et~al.(1976)]{code+76}
A.~D. {Code}, R.~C. {Bless}, J.~{Davis}, and R.~H. {Brown}, \emph{\apj}
  \textbf{203}, 417--434 (1976).

\bibitem[{di Benedetto}(1993)]{dibenedetto93}
G.~P. {di Benedetto}, \emph{\aap} \textbf{270}, 315--334 (1993).

\bibitem[{Jerzykiewicz} and {Molenda-Zakowicz}(2000)]{jerzykiewicz_zakowicz00}
M.~{Jerzykiewicz}, and J.~{Molenda-Zakowicz}, \emph{Acta Astronomica}
  \textbf{50}, 369--380 (2000).

\bibitem[{Katz} et~al.(1998)]{katz+98}
D.~{Katz}, C.~{Soubiran}, R.~{Cayrel}, M.~{Adda}, and R.~{Cautain}, \emph{\aap}
  \textbf{338}, 151--160 (1998).

\bibitem[{Soubiran} et~al.(1998)]{soubiran+98}
C.~{Soubiran}, D.~{Katz}, and R.~{Cayrel}, \emph{\aaps} \textbf{133}, 221--226
  (1998).

\bibitem[{Gregg} et~al.(2004)]{gregg+04}
M.~D. {Gregg}, H.~C. {Ferguson}, D.~{Minniti}, N.~{Tanvir}, and R.~{Catchpole},
  \emph{\aj} \textbf{127}, 1441--1459 (2004).

\end{thebibliography}
\end{document}